\documentclass[pss]{wiley2sp} 
\usepackage{amsmath,amssymb}
\usepackage{bm}             
\usepackage{siunitx}
\DeclareSIUnit\torr{Torr}

\usepackage{graphicx}

\newcommand{\tfg}[1]{\textsubscript{\protect\raisebox{-1pt}{#1}}}

\newcommand{\mtfg}[1]{_{\mathrm{#1}}}			
\newcommand{\mhog}[1]{^{\mathrm{#1}}}

\begin{document}

\title{Nickel oxide-based heterostructures with large band offsets}

\author{%
  Robert Karsthof\textsuperscript{\Ast,\textsf{\bfseries 1,2}},
  Holger von Wenckstern\textsuperscript{\textsf{\bfseries 2}},
  Jes\'{u}s Z\'{u}\~{n}iga-P\'{e}rez\textsuperscript{\textsf{\bfseries 3}},
  Christiane Deparis\textsuperscript{\textsf{\bfseries 3}},
   Marius Grundmann\textsuperscript{\textsf{\bfseries 2,3}}
  }

\mail{e-mail
  \textsf{r.m.karsthof@smn.uio.no}}

\institute{%
  \textsuperscript{1}\,Universitetet i Oslo, Center for Materials Science and Nanotechnology, Gaustadall\'{e}en 23A, Kristen Nygaards hus, 0373 Oslo, Norway\\
  \textsuperscript{2}\,Universit\"{a}t Leipzig, Faculty of Physics and Earth Sciences, Felix Bloch Institute for Solid State Physics, Linn\'{e}str. 5, 04103 Leipzig, Germany\\
  \textsuperscript{3}\, CRHEA-CNRS, rue Bernard Gr\'{e}gory, 06560 Valbonne, France\\
  }


\abstract{\bf%
  We present research results on the electronic transport in heterostructures based on p-type nickel oxide (NiO) with the n-type oxide semiconductors zinc oxide (ZnO) and cadmium oxide (CdO). NiO is a desirable candidate for application in (opto-)electronic devices. However, because of its small electron affinity, heterojunctions with most n-type oxide semiconductors exhibit conduction and valence band offsets at the heterointerface in excess of 1 eV. ZnO/NiO junctions exhibit a so called type-II band alignment, making electron-hole recombination the only process by which a current can vertically flow through the structure. These heterojunctions are nevertheless shown to be of practical use in efficient optoelectronic devices, as exemplified here by our UV-converting transparent solar cells. These devices, although exhibiting high conversion efficiencies, suffer from two light-activated recombination channels connected to the type-II interface, one of which we identify and analyse in more detail here. Furthermore, CdO/NiO contacts were studied – a heterostructure with even larger band offsets such that a type-III band alignment is achieved. This  situation theoretically enables the development of a 2-dimensional electronic system consisting of topologically protected states. We present experiments demonstrating that the CdO/NiO heterostructure indeed hosts a conductive layer absent in both materials when studied separately.
  }

\maketitle   

\section{Introduction}

Due to the abundance of $n$-type oxide semiconductors with wide band gaps on the one hand, but the lack of $p$-type variants of the same materials, the fabrication of all-oxide $pn$ homojunctions for use in transparent electronics is out of reach up to now. In this regard, NiO offers a promising alternative: it has a band gap of around \SI{3.8}{\electronvolt} and is therefore transparent for the human eye; it forms rectifying contacts with a variety of wide-gap $n$-type materials such as ZnO \cite{Ohta2003,Zhang2013a,Karsthof2016}, In$_2$O$_3$ \cite{Wenckstern2015}, Ga$_2$O$_3$ \cite{Schlupp2019} and TiO$_2$ \cite{Ohtsuki2013,Uddin2017}; and it can be doped, both extrinsically (by group-I elements like Li) and intrinsically (by oxygen excess). Consequently, it has been widely used in heterojunction devices where a certain degree of transparency is needed, most prominently as $p$-type electrode in transparent UV detectors and solar cells based on ZnO \cite{Karsthof2016,Zhang2019} and as hole transport layer in perovskite solar cells \cite{Jeng2014,Xu2015,Park2015,Bai2016}. Most recently, it has also found application in gas-sensing devices, both as stand-alone component and in combination with other oxide materials \cite{Tian2016,Deng2017,Gu2017,Nakate2019}. \\
Because of its low electron affinity of \SI{1.46}{\electronvolt} \cite{Wu1997}, NiO has the tendency to form contacts with other materials exhibiting large band offsets in both valence and conduction bands. In most cases, this will lead to a so-called type-II band alignment where two materials share a gap of reduced size at the interface ('staggered gap'). Such a situation is sometimes considered undesirable because it invokes electron-hole recombination at the heterointerface. For that reason, there have only been few attempts to develop a theory that describes the current transport mechanism in a type-II system \cite{Dolega1963,Grundmann2014}). Based on the results cited above, however, it can be stated that this configuration is suitable for efficient bipolar devices. In particular, for solar cells, the high band offsets act as barriers for minority carrier re-injection which overall benefits the device perfomance. In this paper, we will therefore make an attempt to develop a general framework which addresses interface-recombination limited transport in type-II heterojunctions. While the work of Dolega \cite{Dolega1963} focuses on the special case of infinitely fast interface recombination, the theory previously published by us \cite{Grundmann2014} fails to account for ideality factors of well below 2 encountered in the experiments described in this publication. We also show, using the example of the ZnO/NiO contact, that the defect properties of the interface have a pivotal role in determining the parameters of the current-voltage relationship. \\
The second part of the publication is devoted to an electronically even more extreme situation. The conductive transparent material CdO possesses such a high electron affinity that, when in contact with NiO, no interface gap remains -- the band alignment is said to be of 'broken-gap' or type-III character. This configuration holds the promise of enabling the formation of a topological phase at the interface, as will be described further below. Results in the literature on this subject are limited to narrow-gap materials where liquid-He temperatures are necessary to observe the desired effect. Using wide-gap materials instead facilitates its stability also at higher temperatures. Herein, we report for the first time investigations of the lateral electrical transport in CdO/NiO heterocontacts, and we show indirect evidence for a conductive channel forming within the otherwise insulating NiO that is induced by the type-III contact, and that is stable up to room temperature.

\section{Experimental details}

\subsection{ZnO/NiO heterostructures}

For the investigation of transport properties in type-II structures, a \SI{180}{\nano\meter} thick Al-doped ZnO was grown by pulsed laser deposition (PLD) on top of $a$-plane oriented corundum substrates, serving as back contact layer. On top of this, a nominally undoped, \SI{1}{\micro\meter} thick ZnO film was deposited. For both layers, the substrate temperature amounted to \SI{670}{\degreeCelsius}, and an oxygen partial pressure of \SI{0.016}{\milli\bar} was used. Individual pillar-shaped NiO contacts were then fabricated on top by PLD, employing conventional UV photolithography and the lift-off technique for patterning. Before removing the photoresist, the NiO was capped by a \SIrange{3}{4}{\nano\meter} thin Au layer, fabricated by DC magnetron sputtering, which is needed to ensure uniform current extraction across the contact area. Due to the use of the temperature-sensitive photoresist, NiO growth was conducted at room temperature (no intentional substrate heating). We chose a comparably high oxygen partial pressure of \SI{0.1}{\milli\bar} which is known to lead to high acceptor (Ni vacancy) densities in NiO, therefore rendering the film semiconducting. Semitransparent ZnO/PtO$_x$ Schottky diodes were fabricated by reactive dc magnetron sputtering from a metallic Pt target in O\tfg{2} atmosphere, and patterning of these structures was done in the same way as for the NiO contact. Further details on the PLD system and our fabrication processes can be found in Refs.~\cite{Lajn2009,Mueller2014,Karsthof2016,Karsthof2016a}. \\
Characterization of the strucures by current density-voltage ($j$-$V$) measurements was conducted using a S\"{U}SS wafer prober with tungsten needles and an Agilent 4155C precision semiconductor parameter analyzer. On the basis of these measurements, individual contacts were selected for further characterization. Those samples were then mounted onto transistor sockets, and the selected pillars were contacted by Au wire bonding using silver epoxy resin. Capacitance-voltage ($C$-$V$) measurements were conducted with the help of an AGILENT Precision Semiconductor Parameter Analyzer 4294A. For characterization under UV illumination, we used a high-power UV LED (emission centered around \SI{365}{\nano\meter}, total optical power density \SI{160}{\milli\watt\per\square\centi\meter}).

\subsection{CdO/NiO heterostructures}

The CdO (100) layers reported here were grown by plasma-assisted molecular beam epitaxy on top of (100) oriented MgO substrates at a Cd effusion cell temperature of \SI{400}{\degreeCelsius} and an oxygen flow of \SI{0.4}{\cubic\centi\meter\per\minute}, giving rise to a residual pressure during growth of about \SI{e-5}{\torr}. For the CdO/NiO heterostructures, uniform, \SI{50}{\nano\meter} thick NiO thin films were grown by PLD on top of the CdO, using a substrate temperature of \SI{320}{\degreeCelsius} and an oxygen partial pressure of \SI{e-2}{\milli\bar}. This was done to obtain insulating NiO films so as to eliminate as many conduction channels in the structures as possbile, facilitating the identification of (potentially weak) additional conductivities induced by making the CdO/NiO contact.\\
The samples were characterized by Hall effect and magnetoconductivity measurements in a Physical Properties Measurement System (Quantum Design).


\section{Type-II band alignment: ZnO/NiO}

\begin{figure}
	\centering
	\includegraphics[width=0.9\columnwidth]{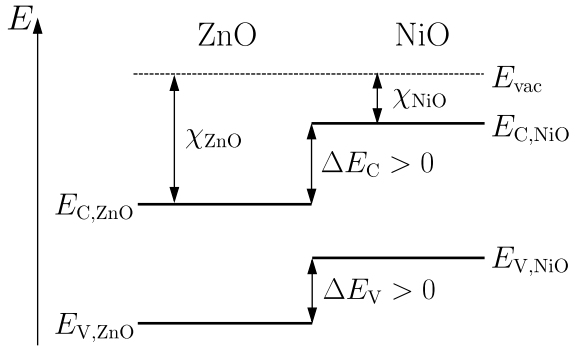}
	\caption{Schematic representation of the type-II band alignment between ZnO and NiO.}
	\label{fig:type2_alignment}
\end{figure}

The most thoroughly investigated heterostructure including NiO is with ZnO as $n$-type partner. ZnO has a wide band gap of around \SI{3.3}{\electronvolt}, can be deposited and patterned easily by technologically relevant methods, and is earth-abundant and non-toxic. Because of its low electron affinity, NiO forms a type-II band alignment with ZnO (electron affinity above \SI{4}{\electronvolt} \cite{Jacobi1984}). In such an arrangement, valence and conduction bands exhibit offsets $\Delta E\mtfg{C,V}$ with the same sign (Fig.~\ref{fig:type2_alignment}), sometimes also termed \textit{staggered gap alignment}. The magnitudes of the respective offsets have been determined by various researchers and are summarized in Table~\ref{tab:NiO-ZnO_offsets}. Although these values vary significantly, the strong type-II alignment with $\Delta E\mtfg{C,V} > \SI{1}{\electronvolt}$ is supported by all available experimental results.

\begin{table}
  \caption{Literature values for the valence and conduction band offsets for the ZnO/NiO heterocontact.}
  \centering
  \begin{tabular}[]{SSll}
    \hline
    {$\Delta E\mtfg{V}$ (\si{\electronvolt})} &  {$\Delta E\mtfg{C}$ (\si{\electronvolt})} & remarks & reference \\
    \hline
    	2.9		&	3.2	 	& 	&		\cite{Ishida2006}			\\
    	2.60 	& 	2.93 	&  	&		\cite{Deng2009}				\\
	   	1.47	& 	1.80	&	&		\cite{Yang2011}  			\\
    	1.6		& 	2.00	&	non-polar ZnO	&	\cite{Ma2013}  \\
    	1.3		&	1.72	&	polar ZnO		&	\cite{Ma2013}	\\	
    	1.6		&	2.0		&	&\cite{Kawade2014}					\\
    	2.04	&	2.54	&	\textbf{our work}	&				 \\
    \hline
  \end{tabular}
  \label{tab:NiO-ZnO_offsets}
\end{table}

\subsection{Characterization of the electrical transport: the role of interface defects}
One can easily imagine that strong band offsets impose injection barriers for charge carriers: in the case discussed in this paper, electron currents from ZnO into NiO, and hole currents from NiO to ZnO, respectively, will be strongly suppressed. A current can only flow by means of electron-hole recombination at the interface, which is usually supported by interfacial electronic defect species acting as recombination centers. The behavior of this contact, e.g. its characteristics under electric biases or illumination, and therefore also its performance in active electronic devices, are strongly dependent on the detailed interface properties. More specifically, the voltage dependence of the current flowing in such a situation (the current density-voltage or $j$-$V$ characteristic of the contact) can be expected to depend strongly on the recombination rate associated with the interfacial defect states, i.e., their density, carrier capture and emission rates etc. Finding an analytical expression describing the $j$-$V$ curve of such a contact which captures all these parameters under generalized conditions is tremedously difficult. Dolega \cite{Dolega1963} has calculated an expression under the assumption of infinitely fast recombination at the interface, implying that it is the flow of free carriers to the interface, not the interface recombination iself, that limits the total current. This is, however, not necessarily applicable to many relevant cases, where differences in interface properties are reflected strongly in measured $j$-$V$ characteristics. This is exemplified by the following experiment. Two thin film heterostructures were produced consisting of NiO on top of ZnO.
For the first sample, the initial \num{5000} pulses of the NiO deposition were conducted with a circular shadow mask inserted into the center of the plasma plume. This leads to significantly reduced growth rates on the one hand, but prevents high-energy particles from directly reaching and damaging the surface of the growing film on the other hand, thereby suppressing the formation of impact-induced defects \cite{Schubert2015}. The technique is also referred to as \textit{eclipse-PLD}. After the deposition of a \SI{2}{\nano\meter} thin layer using this method, the rest of the NiO film (approximately \SI{100}{\nano\meter}, corresponding to \num{30000} laser pulses) was deposited without the aperture. For the second sample, a NiO layer of the same thickness as in the first sample was deposited directly on top of the ZnO without modifications ('standard NiO'), serving as a reference.

\begin{figure}
	\centering
	\includegraphics[width=\columnwidth]{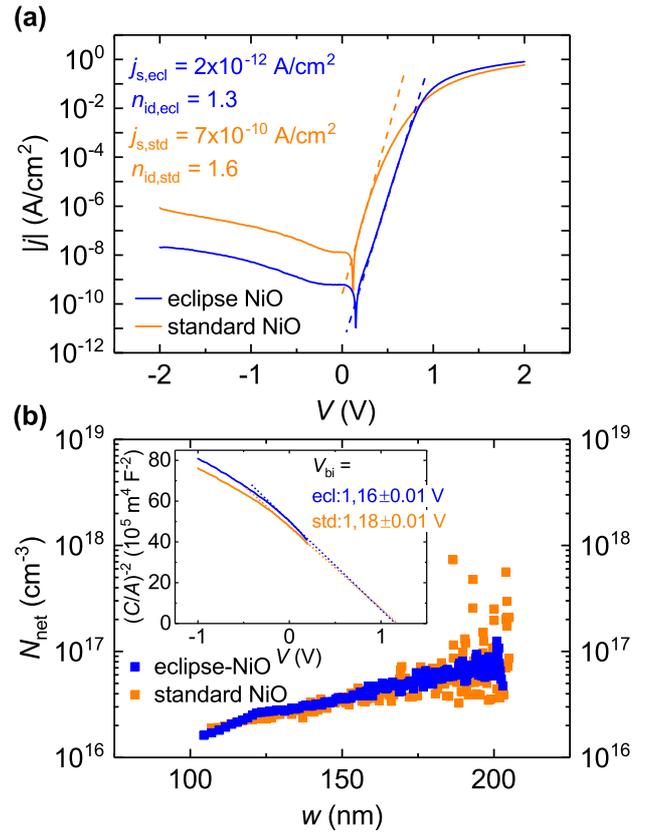}
	\caption{(a) Current-voltage characteristics for ZnO/NiO heterocontacts from interface manipulation experiment, and fits in the forward region acc. to (\ref{eq:diode}). (b) Doping profiles for the same contacts, obtained from $C$-$V$ measurements; inset shows corresponding Mott-Schottky plots and extrapolated built-in voltages.}
	\label{fig:NiO-ZnO_IF}
\end{figure}

The characteristics are shown in Fig.~\ref{fig:NiO-ZnO_IF}a as absolute current densities in a semilogarithmic plot. The figure also includes fits to the curves in the forward voltage region according to the general diode relation 

\begin{equation}
	j(V) = j\mtfg{s}\left[\exp\left(\frac{\beta V}{n\mtfg{id}}\right)-1\right]
	\label{eq:diode}
\end{equation}

with $n\mtfg{id}$ and $j\mtfg{s}$ being the ideality factor and saturation current density of the diode, respectively, and $\beta = \frac{e}{k\mtfg{B}T}$. Values for $j\mtfg{s}$ and $n\mtfg{id}$ are given in the figure. It can be seen that the interface quality clearly has an inpact on the diode behavior. The eclipse-PLD sample exhibits a $j\mtfg{s}$ two orders of magnitude lower than the reference sample, and $n\mtfg{id}$ much closer to 1 than in the reference. In our previous attempt to develop a theory for recombination-limited transport in type-II structures \cite{Grundmann2014}, we derived the ideality factor of such a contact to be always around 2, only with minor variations depending on the doping symmetry and the interface recombination rate. The result here is in clear contradiction to that prediction. \\
What follows is a more crude and general derivation, lacking quantitative prediction power on the one hand, but being able to reproduce the results presented here in a qualitative manner on the other, in contrast to Ref.~\cite{Grundmann2014}. To develop a model for the current-voltage relationship that considers the interface recombination itself to be the limiting process, and not the transport \textit{to} the interface, as done by \cite{Dolega1963}, the following expressions for the current densities on both sides of the junction can be used:

\begin{equation}
	j\mtfg{p} = e S\mtfg{p} \delta p\mtfg{if}, \quad j\mtfg{n} = e S\mtfg{n} \delta n\mtfg{if}
\end{equation}

with interfacial excess carrier densities $\delta p$, $\delta n$ and interface recombination velocities $S\mtfg{p,n}$ for holes on the $p$-side and electrons on the $n$-side, respectively. Using the Boltzmann approximation, assuming that the donor density $N\mtfg{D}$ on the $n$-side is considerably lower than the acceptor density $N\mtfg{A}$ on the $p$-side (a situation that applies to these diodes, as will be shown further below), one can show that the $j$-$V$ relation takes the form

\begin{equation}
	j(V) = e S\mtfg{n}N\tfg{D} \exp\left[-\beta V\mtfg{bi}\right] \left(\exp\left[\beta V\right]-1\right),
\end{equation}

$V\mtfg{bi}$ being the built-in voltage of the contact. Comparing this with (\ref{eq:diode}), we identify the parameters $j\mtfg{s}$ and $n\mtfg{id}$ as

\begin{equation}
	j\mtfg{s} = e S\mtfg{n} N\mtfg{D} \exp\left[-\beta V\mtfg{bi}\right] \quad \text{and} \quad n\mtfg{id} = 1.
	\label{eq:diode_param_recombination_asym}
\end{equation}

The recombination velocity $S\mtfg{n}$, on the other hand, can be described by ${S\mtfg{n} = \sigma\mtfg{c,n} v\mtfg{th,n} N\mtfg{if}}$ with $\sigma\mtfg{c,n}$ being the defects' capture cross section for electrons, $v\mtfg{n,th}$ the electron thermal velocity in the $n$-type semiconductor, and $N\mtfg{if}$ the interface defect density. Collecting all factors, we have

\begin{equation}
 j\mtfg{s} = e N\mtfg{D} \sigma\mtfg{c,n} v\mtfg{th,n} N\mtfg{if} \exp\left[-\beta V\mtfg{bi}\right].
 \label{eq:js_IF_rec}
\end{equation}

$v\mtfg{th,n}$ and $\sigma\mtfg{c,n}$ are considered here to be independent of the interface quality. The changes of $j\mtfg{s}$ observed in the above experiment must therefore be related to either the ZnO doping, $N\mtfg{D}$, the interface defect density $N\mtfg{if}$ or the built-in voltage $V\mtfg{bi}$. To clarify this, capacitance-voltage ($C$-$V$) measurements were taken at room temperature and an AC frequency of \SI{10}{\kilo\hertz}. The results are shown in Fig.~\ref{fig:NiO-ZnO_IF}b in the form of doping profiles. Both samples have comparable net doping levels of some \SI{e16}{\per\cubic\centi\meter}, which is two orders of magnitude lower than the NiO doping \cite{Karsthof2019}, justifying the assumption of asymmetric doping levels made above. It should be added that the eclipse-PLD technique is known to cause slightly less compact thin films due to the reduced particle surface energies \cite{Schubert2015}, but since the NiO doping is determined mostly by the oxygen partial pressure during growth, the doping concentration of the thin eclipse layer can be considered to be similar to the one of the conventionally deposited film. 

The inset of Fig.~\ref{fig:NiO-ZnO_IF}b displays Mott-Schottky plots ($C\mhog{-2}$ vs. $V$) from the same measurements, which have been extrapolated to obtain the built-in voltages which are very similar for both samples. The interface quality therefore does not have any impact on the total band bending. As a consequence, a saturation current density of the eclipse-NiO sample reduced by about a factor of 350 in comparison to the reference sample can only be related to an interface defect density reduced by the same extent. Apart from defect states that may form in the mere presence of a heterointerface (which, in itself, is nothing but a 2-dimensional structural defect naturally producing a certain population of in-gap states), structural defects may also be introduced during growth of the heterostructure. The surface of one semiconductor (more specifically: the uppermost few atomic layers) exposed to an incoming particle flux during deposition of a second semiconductor may be damaged upon impact of these particles. Eclipse-PLD is a method specifically designed to suppress such damaging, and has been shown to be successful in doing so, e.g. for the deposition of a TCO on top of organic semiconductors \cite{Schubert2015}. In the present case, it seems to be appropriate to prevent damaging of the ZnO surface as well. Furthermore, this experiment unambiguously demonstrates that the interface recombination itself, and not the transport of charge carriers to the interface, is the critical process in this particular heterostructure. Moreover, ideality factors deviating strongly from 1 are, in many cases, indicative of laterally varying diode properties within the contact area \cite{Werner1991}. The fact that $n\mtfg{id}$ is closer to 1 for the eclipse-PLD sample can therefore be interpreted as being due to better homogeneity -- another indication for the critical role of the interface in determining the parameters of the electronic transport in these particular type-II heterostructures.

%


\subsection{Transparent UV-active solar cells based on ZnO/NiO}
\begin{figure}
	\centering
	\includegraphics[width=\columnwidth]{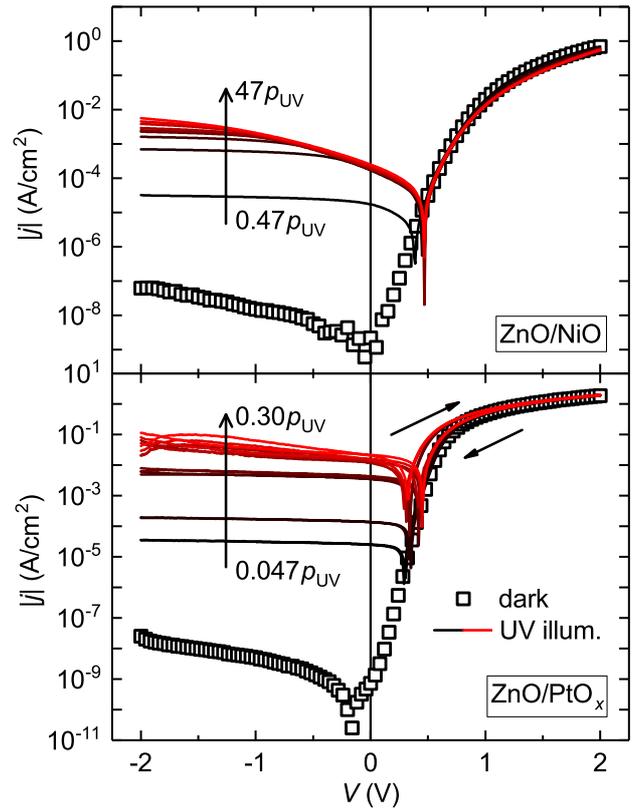}
	\caption{Current-density voltage relations of a ZnO/eclipse-NiO (top panel) and ZnO/PtO$_x$ Schottky (bottom panel) solar cell under UV illumination for different intensities, given relative to the solar terrestrial UV power density. Dark characteristics are also shown.}
	\label{fig:ZnO-NiO_UV_illu_jV}
\end{figure}

\begin{figure}
	\centering
	\includegraphics[width=\columnwidth]{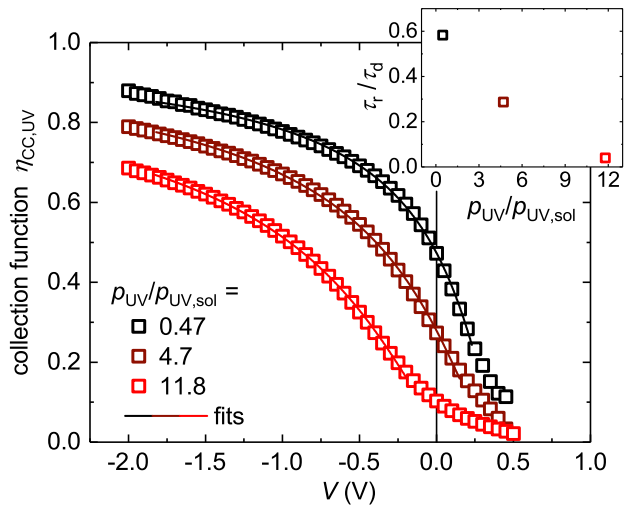}
	\caption{Voltage dependence of the carrier collection efficiency of a ZnO/eclipse-NiO solar cell under monochromatic UV excitation for different illumination intensity, and fits acc. to (\ref{eq:coll_eff_SCR}).}
	\label{fig:ZnO-NiO_etaCCUV}
\end{figure}

Based on ZnO/NiO heterojunctions, we already demonstrated a visible light-transparent, UV-converting solar cell with a conversion efficiency of 3.1\% with respect to the UV part of the solar spectrum \cite{Karsthof2016}. The functionality of these oxide interface devices relies on the absorption of UV light by ZnO (photon energies $>\SI{3.2}{\electronvolt}$) and the subsequent separation of the generated electron-hole pairs at the interface towards the NiO layer. The high hole density in the NiO ensures that the ZnO layer with its lower doping accommodates the main part of the space charge width and band bending, such that as many generated carriers as possible can be collected. The band offsets between the two materials produce additional potential barriers for back-diffusion of photo-generated carriers once they have crossed the interface.\\
The critical contribution to the photocurrent is respresented by the holes (minorities) in the space charge region flowing towards the interface; their lifetime limits the collection efficiency. We have analyzed the electric loss processes in these solar cells and concluded that white-light (i.e. solar light) illumination increases the interface recombination current, which reduces photovoltaic perfomance in terms of all three parameters open-circuit voltage $V\mtfg{oc}$, short-circuit current density $j\mtfg{sc}$ and fill factor $f\mtfg{f}$ \cite{Karsthof2016a}. This effect that, so far, could not be tied to a specific photon threshold energy, increases the probability for an interface-crossing hole to recombine on the one hand, and enables higher interface recombination velocities for injected carriers on the other hand. The latter is expressed in the form of a so-called \textit{cross-over}, i.e. an intersection of the $j$-$V$ characteristics in dark and under illumination with simulated sunlight. This intersection occurs close to $V\mtfg{oc}$, implying that the open-circuit state is adversely affected. \\
This white-light induced process is, however, not the only loss mechanism in ZnO/NiO solar cells that is triggered by illumination. We have tested our devices under monochromatic ultraviolet excitation, using a high-power UV LED and employing different neutral density filters to adjust the illumination intensity between \SI{5}{\percent} and \SI{4700}{\percent} of the UV power density contained in the terrestrial sunlight (AM1.5G). For this study we investigated the eclipse-NiO/ZnO device described in the paragraph above, which was done to ensure a high quality interface with the lowest possible density of structural defects. The resulting illumination-dependent $j$-$V$ characteristics are shown in Fig.~\ref{ZnO-NiO_UV_illu_jV} (top panel). Apparently, the ZnO/NiO cell is not able to collect the surplus of charge carriers generated at a relative power density higher than approximately \SI{100}{\percent} without a significant reverse voltage being applied. The bottom panel of Fig.~\ref{fig:ZnO-NiO_UV_illu_jV} displays analogous measurements on a semitransparent Schottky contact fabricated on top of the same ZnO material, using oxidized platinum (PtO$_x$) as transparent rectifying contact \cite{Lajn2009,Mueller2014}. It is evident that this structure is not only giving a higher photocurrent density for the same neutral density filter values, the voltage dependence of its photocurrent density is also much weaker. The loss process responsible for the strong voltage dependence must therefore be connected to the usage of NiO as front contact. It is also apparent from the data that cross-over of the dark and illuminated characteristics is absent for the ZnO/NiO cell, which indicates that there is no connection to the white light-enhanced interfacial recombination current identified in Ref.~\cite{Karsthof2016a}. For the Schottky reference structure, the forward currents are slightly enhanced for voltage sweeps from negative to positive bias, leading to a hysteresis of the characteristics between the two sweep directions. Such an effect has been found in (Mg,Zn)O-based structures \cite{Zhang2011,Zhang2013} and has been attributed to the trapping and detrapping of photogenerated holes at the metal-semiconductor interface. It is likely that the situation is similar in the present case. No hysteresis has been found in our ZnO/NiO devices.\\
In order to analyze the UV-enhanced loss process further, we have employed the method described in Ref.~\cite{Karsthof2016a} to calculate and fit the voltage dependence of the carrier collection efficiency under UV illumination, $\eta\mtfg{CC,UV}$, using the $j$-$V$ characteristics shown in Fig.~\ref{fig:ZnO-NiO_UV_illu_jV}. The procedure has been applied to three pairs of characteristics, pairwise similar with respect to $p\mtfg{UV}$, to obtain $\eta\mtfg{CC,UV}$ for weak, medium and strong UV illumination conditions (Fig.~\ref{fig:ZnO-NiO_etaCCUV}). A significant decrease of $\eta\mtfg{CC,UV}$ and increasingly convex curvature of the collection function in the photovoltaic quadrant with intensifying UV is observed. For negative voltages, the data can be well fitted with a model considering recombination within the space charge region \cite{Misiakos1988}: 

\begin{equation}
	\eta\mtfg{CC}(V) = \frac{\tau\mtfg{r}}{\tau\mtfg{d}}\left(1-\frac{V}{V_0}\right)\left(1-\exp\left[-\frac{\tau\mtfg{d}}{\tau\mtfg{r}\left(1-\frac{V}{V_0}\right)}\right]\right)
	\label{eq:coll_eff_SCR}
\end{equation}

with $\tau\mtfg{r}$ and $\tau\mtfg{d}$ recombination lifetime and drift time for minority carriers in the depletion region, respectively, and $V_0$ a characteristic voltage for which the collection probability goes to zero. The inset in Fig.~\ref{fig:ZnO-NiO_etaCCUV} shows how the ratio of recombination and drift time decreases with intensifying UV illumination, underlining the light-activated character of the process. Recombination within the space charge region is usually caused by point defects acting as recombination centers. Because the loss mechanism is not present for the Schottky reference structure, the defect in question must be specifically related to the usage of NiO. Ni as an impurity in ZnO has been investigated, e.g., by Schmidt \textit{et al.} \cite{Schmidt2011} where the authors show Ni incorporation to lead to the formation of an electronically defect having two detectable charge transistion within the ZnO band gap. The deeper of these two can be (de-)populated by UV photons. Based on other available works on Ni in Zn chalcogenides, the substitutional defect Ni\tfg{Zn} seems most plausible as a candidate.

\begin{figure}
	\centering
	\includegraphics[width=\columnwidth]{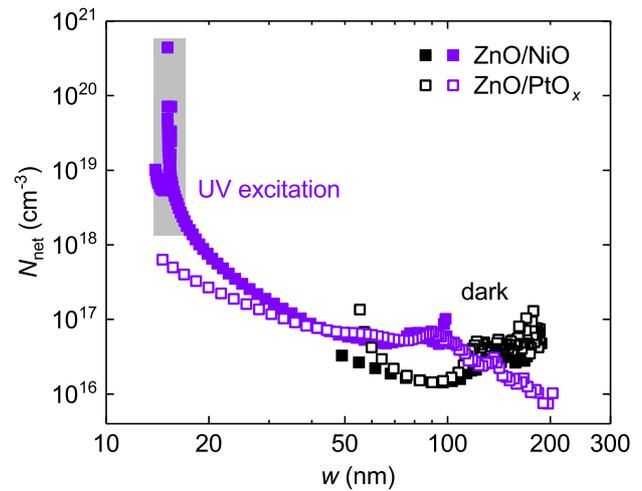}
	\caption{Doping profiles of ZnO-based solar cells with an eclipse-NiO- and a Schottky front contact, in dark and under illumination with a UV LED ($p\mtfg{opt,UV}=\SI{16}{\milli\watt\per\square\centi\meter}$), collected at a measurement frequency of \SI{4}{\kilo\hertz}.}
	\label{fig:photo-CV_UV}
\end{figure}

Fig.~\ref{fig:photo-CV_UV} shows doping profiles, obtained by $C$-$V$ measurements, from an NiO-based and a Schottky contact solar cell in the dark state and illuminated by the UV LED (a neutral density filter was used to reduce the power density to \SI{16}{\milli\watt\per\square\centi\meter}). While for the Ni-free cell, the apparent net doping density increases only moderately without significant changes to the overall profile shape, the NiO-based structure exhibits a pronounced peak in $N\mtfg{net}$ close to the interface towards the NiO. Because this feature is absent in the PtO$_x$ control sample, we attribute it to the aforementioned defect Ni\tfg{Zn} acting as UV-activated recombination center. Ni incorporation is possible by in-diffusion during NiO growth, although it shall be mentioned that (i) the growth itself was conducted at room temperature and only the sample contacting involved a slightly elevated temperature of \SI{90}{\degreeCelsius}, and (ii) that an implantation of Ni can probably be excluded because the sample investigated here has an eclipse-NiO contact. Ni therefore seems to have high diffusivity in ZnO, also at moderate temperatures.\\
Although the loss process identified here dominates only at comparatively high power densities that are most likely not relevant for real applications, it is also limiting the cell performance under normal AM1.5G conditions to a considerable extent. Using the data in the inset of Fig.~\ref{fig:ZnO-NiO_etaCCUV}, it is evident that for $p\mtfg{UV}/p\mtfg{UV,sol} = 1$ the parameter $\tau\mtfg{r}/\tau\mtfg{d}$ is around \num{0.5}. With the aid of (\ref{eq:coll_eff_SCR}) and considering short-circuit conditions ($V=\SI{0}{\volt}$), this translates to a fraction of \num{0.57} of the photogenerated minorities being lost to recombination in the space charge layer. This demonstrates that the impact of the UV-activated process should not be underestimated.

\section{Type-III band alignment: CdO/NiO}

\begin{figure}
	\centering
	\includegraphics[width=\columnwidth]{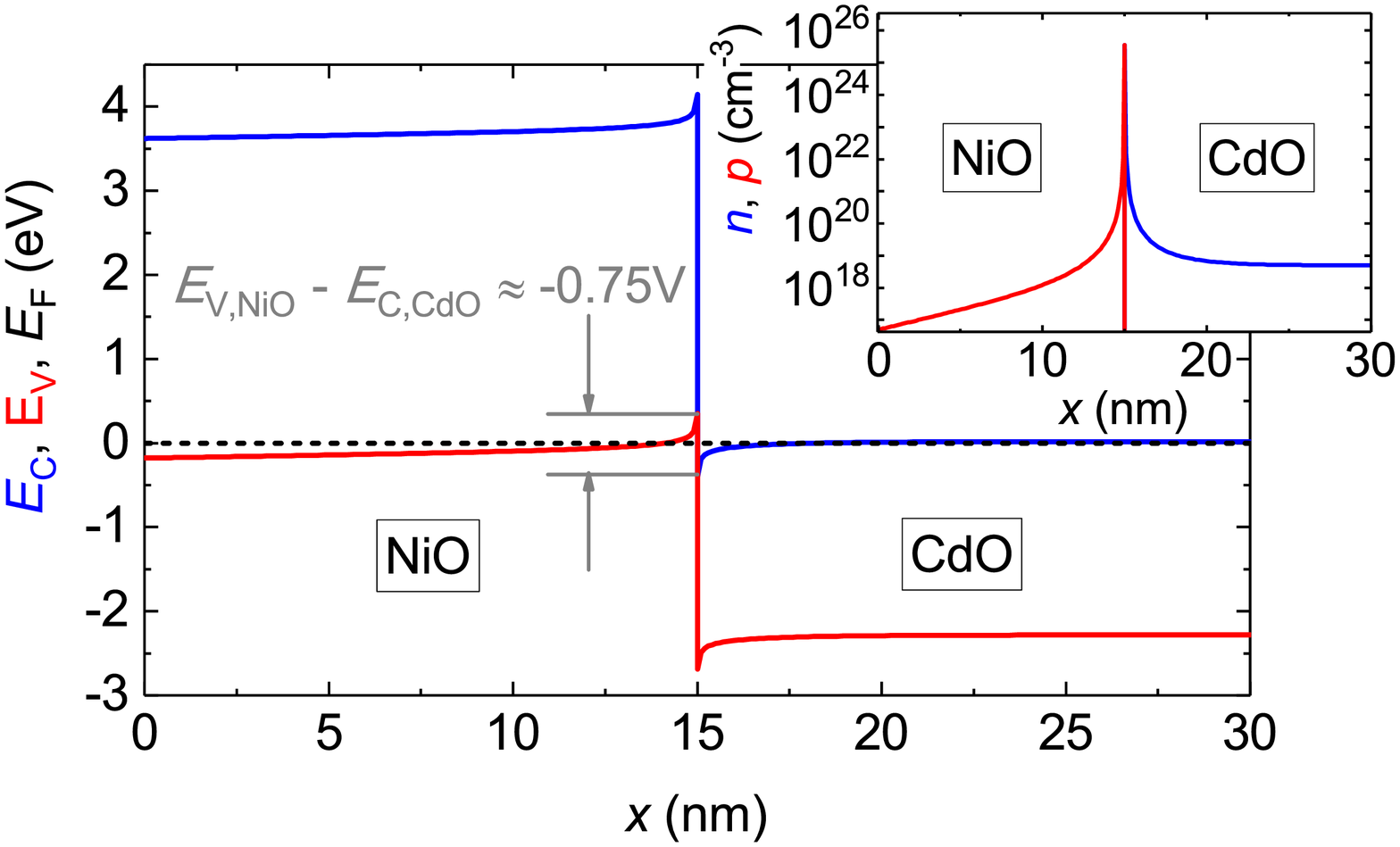}
	\caption{Simulated band diagram of a thin NiO/CdO heterostructure in thermodynamical equilibrium. The enhancement of the carrier densities in the interfacial region is shown in the inset.}
	\label{fig:NiO-CdO_band_alignment}
\end{figure}

Cadmium oxide (CdO) belongs to the materials showing high electrical conductivity and optical transparency simultaneously (transparent conductive material, TCM), and was one material system amongst a few others for which this property was reported for the very first time, namely by Karl B\"{a}deker in 1907 \cite{Baedeker1907,Grundmann2015}. Its strong propensity for $n$-type conductivity stems from its low-lying, highly dispersive conduction band (electron affinity \SIrange{5.8}{6.0}{\electronvolt} \cite{Francis2015}). Combined with the high-lying valence band of NiO, a CdO/NiO heterostructure opens the possibility of the formation of a type-III (or 'broken-gap') band alignment. The valence band states of NiO and the conduction band states of CdO are similar in energy, with an expected overlap of the two bands of the order of \SI{1}{\electronvolt} \cite{Francis2015}. This can enable coupling between these states by tunneling at the interface. Because the CdO CB is mostly Cd 5s derived, and the NiO VB is a hybrid mixture of Ni 3d and O 2p states, a set of bands with mixed parity is expected to evolve at the interface, separated by an energy gap. Similar effects have already been demonstrated to occur in CdTe/HgCdTe heterostructure quantum wells \cite{Konig2007} and InAs/GaSb heterostructures \cite{Knez2011}, with evidence supporting transport through so-called \textit{helical edge states} with quantized conductivity, indicative of the opening of a hybrid gap of the order of a few \si{\milli\volt} in both cases. In comparison to the materials previously investigated, both CdO and NiO have larger band gaps (\SI{2.3}{\electronvolt} and \SI{3.8}{\electronvolt}, respectively) which, on the one hand, may give rise to a larger hybridization gap and thereby stability of helical edge state transport at much higher temperatures. On the other hand, the transparency of both materials for a wide range of photon energies may enable studying interface properties by optical methods.\\
We have used the tool SCAPS-1D \cite{Niemegeers1996,Burgelman2000} to numerically calculate the band diagram of the NiO/CdO heterostucture and the charge carrier density distribution within it in thermodynamic equilibrium. The relevant material parameters used for the calculation are given in Table~\ref{tab:sim_param}. The film thicknesses were set to be \SI{15}{ \nano\meter} each. The doping level values used for the calculation were chosen as typically measured for our CdO films produced for this study; for NiO we assumed a doping level in a range where the NiO film is usually not conductive anymore, like it is realized in the films presented further below. For both materials we assumed a spatially constant charge carrier density. The result of the simulation is shown in Fig.~\ref{fig:NiO-CdO_band_alignment}. The broken gap-alignment is clearly demonstrated by a negative offset between the CdO conduction band and the NiO valence band being as large as \SI{-0.75}{\electronvolt}. It is also shown that the carrier densities on both sides of the interface are significantly enhanced within a few \si{\nano\meter}, and that CdO and NiO show electron and hole enhancement, respectively. It should be noted that the presence of an intrinsic accumulation layer on the CdO surface, as will be shown to exist further below, is not accounted for in this simulation. 

\begin{table}
	\caption{Relevant material parameters band gap energy, electron affinity and acceptor/ donor densities used for the SCAPS-1D simulation.}
	\label{tab:sim_param}
	\centering
	\begin{tabular}{ccc}
	\hline
	parameter	&	{NiO}	&	{CdO}	 \\
	\hline
	$E\mtfg{g}$ ($\si{\electronvolt}$)&	3.8	\cite{Adler1970}	&	2.3	\cite{King2010} \\
	$\chi$ ($\si{\electronvolt}$) &	1.46 \cite{Wu1997}	&	6.0	\cite{Speaks2010} \\
	$N\mtfg{A/D}$ ($\si{\per\cubic\centi\meter}$)	&	\num{5e16} 	&	\num{5e18}		\\
	\hline
	\end{tabular}
\end{table}

An anti-crossing behavior of extended Cd, O and Ni bands and localized Ni $3d$ levels has already been demonstrated within the bulk of Ni$\mtfg{x}$Cd$\mtfg{1-x}$O alloys \cite{Francis2015,Francis2017,Liu2019}. This can be viewed as a first proof-of-principle for the idea of making a type-III heterostructure with the same materials exhibiting topologically protected states at the interface. In this work, we fabricated such structures for the first time and investigated their electrical transport properties via temperature-dependent Hall effect and magnetoresistance measurements. To unambiguously detect conductive 2-dimensional states in a CdO/NiO heterostructure by such methods, the behaviour of the individual layers must be known first. On the one hand, the NiO films fabricated for this section exhibit electrical conductivity which is orders of magnitude lower than that of CdO, and moreover do not produce a Hall voltage when exposed to a magnetic field. CdO, on the other hand, is highly conductive. Therefore, we first focus on its electrical characterization in the next section before analyzing properties of CdO/NiO heterostructures.

\subsection{Electronic transport in single CdO layers}

It is well established that CdO, like several other $n$-type oxide semiconductors, shows electron accumulation on its surface \cite{King2011}, which can be attributed to the fact that all defects with low formation energy in this material system, intrinsic bulk and surface as well as H-derived defects, are donors \cite{King2009}. The result is a downward band bending towards the surface which can amount to up to \SI{0.6}{\electronvolt} \cite{Francis2017}. It is therefore important to distinguish electrical transport associated with this 'intrinsic' surface accumulation from one that arises due to the possible formation of a type-III heterointerface, which by itself should accomodate enhancement layers.\\

\begin{figure}
	\centering
	\includegraphics[width=\columnwidth]{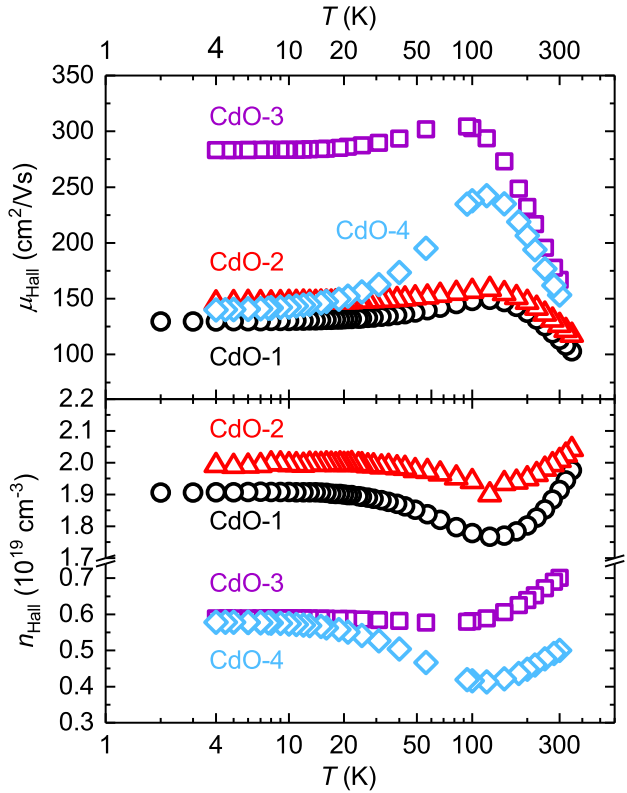}
	\caption{Temperature-dependent electron concentrations and (Hall) mobilities for three CdO films on MgO substrates.}
	\label{fig:CdO-TdH}
\end{figure}

Fig.~\ref{fig:CdO-TdH} shows the temperature dependence of the electron concentration and the Hall mobility obtained on four CdO samples, the details of which will be given further below. Their qualitative behavior is similar: after an initial decrease (increase) of $n\mtfg{Hall}$ ($\mu\mtfg{Hall}$) while $T$ decreases from room temperature, it reaches a minimum (maximum) at $T\approx \SI{100}{\kelvin}$, then increases (decreases) again and finally saturates at low temperatures. This behaviour is typical for samples accomodating several charge carrier systems, of which (at least) one is degenerate and one is not. At low enough temperatures, only the degenererate system can contribute to conductivity, giving temperature-independent $n\mtfg{Hall}$ and $\mu\mtfg{Hall}$. As the temperature increases, the non-degenerate system begins to contribute until at sufficiently high temperatures it dominates the conduction process. Additionally, if the degenerate electrons are not a bulk (3D) but a surface or interface (2D) system, this can lead to extremal values in Hall concentration and mobility \cite{Look1997}, like it is observed in this experiment. The temperature at which these extremal values occur depends on the ratio of the conductances of the two systems. 

To determine the location of the 2D electronic system present in our CdO films, we can compare the four samples with respect to their specifications which we will now lay out. CdO-1 was grown under the standard conditions given above, and serves as a reference. CdO-2 consists of a standard layer grown on top of a low-temperature buffer layer which is employed to reduce the possible influence of electrically active defects at the substrate-film interface. These are known to affect the transport properties of, e.g., GaN layers \cite{Xu2000} and ZnO \cite{Tampo2004} layers on sapphire substrates through the in-diffusion of Al into the film. As can be seen, the inclusion of such a buffer layer does not lead to significantly different electrical properties. Therefore, the presence of a possible degenerate layer at the film/substrate interface can be excluded.

Sample CdO-3 is nominally identical to CdO-1, except it was deposited in a longer growth process, which lead to larger film thickness (\SI{270}{\nano\meter} instead of \SI{115}{\nano\meter}). CdO-3 shows better crystalline quality than CdO-1 in terms of the (200) reflex rocking curve broadening (FWHM see Table~\ref{tab:CdO-samples}). This is consistent with earlier observations on CdO (100) layers grown by MOCVD on $r$-plane sapphire, where a continuous crystal quality improvement with CdO thickness was observed \cite{Zuniga-Perez2004}.
Atomic force microscopy measurements also showed that CdO-3 indeed seems to consist of larger crystallites, which is reflected by its higher rms surface roughness value. In CdO-3, at room temperature, $n\mtfg{Hall}$ is reduced by a factor of three, and $\mu\mtfg{Hall}$ at low temperatures is twice as large with respect to CdO-1. The temperature at which the extremum in $n\tfg{Hall}$ and $\mu\mtfg{Hall}$ appear is also shifting to lower temperatures, which indicates a stronger dominance of bulk over surface conductivity which may be attributed to the increased thickness of CdO-3. The fourth sample, CdO-4, is comparable to CdO-3 with respect to the CdO layer, but it has additionally been capped with a \SI{10}{\nano\meter} thick MgO film \textit{in situ}, i.e. without breaking the vacuum after CdO deposition. It can be seen that its carrier density, including temperature behavior, is similar to that of CdO-3 while the electron mobility at low $T$ is strongly reduced.  

\begin{table*}
	\caption{CdO sample details: film thickness, rms roughness, FWHM of (200) reflex rocking curve, and sheet carrier concentration at low temperatures.}
	\centering
	\begin{tabular}{ccSSSS}
	\hline
	sample & remarks & {$d\mtfg{film}$ (\si{\nano\meter})} & {$r\mtfg{rms}$ (\si{\nano\meter})} & {$\Delta\omega\tfg{(200)}$ (\si{\degree})} & {$n\mtfg{s}(\SI{4}{\kelvin})$ (\SI{e14}{\per\square\centi\meter})} \\
	\hline
	CdO-1 & reference & \num{115} & \num{0.8} & \num{0.45} & \num{2.19}  \\
	CdO-2 & w/ buffer layer & \num{109} & \num{1.1} & & \num{2.17}  \\
	CdO-3 & long growth run & \num{269} & \num{1.6} & \num{0.37} & \num{1.58}  \\
	CdO-4 & w/ MgO capping & \num{249} & \num{0.9} & \num{0.37} & \num{1.43}  \\
	\hline
	\end{tabular}
	\label{tab:CdO-samples}
\end{table*}

The presence of a quasi 2-dimensional electronic system at the free surface of CdO has already been unambiguously demonstrated \cite{Piper2008,King2010}. Electrons are strongly confined to a region in the range of approximately \SI{2.5}{\nano\meter} below the CdO surface, which leads to the quantization of the conductive states into two non-parabolic subbands. This surface system determines $\mu\mtfg{Hall}$ and $n\mtfg{Hall}$ at low temperatures. The differences and similarities between the samples with lower crystalline quality (i.e. CdO-1 and 2) and the one with higher quality (CdO-3) can be understood in the following way: the Fermi level at the CdO surface is pinned to a rather universal value of approximately \SI{0.4}{\electronvolt} above the conduction band minimum \cite{King2010}. This produces a (more or less) fixed surface electron density, as measured by Hall effect at low temperatures (compare the sheet densities $n\mtfg{s}$ in Table~\ref{tab:CdO-samples}). The position of the bulk Fermi level depends on the doping, and the lower the bulk electron density, the larger the difference between bulk and surface Fermi level -- therefore, the deeper the 'surface quantum well'. This means that a lower bulk doping level, as seen in sample CdO-3, leads to a stronger electronic confinement at the surface, which explains the higher mobility exhibited by the surface system in CdO-3. Apparently, depositing a thin passivating MgO layer does not alter the sheet density of the surface system, as exemplified by the comparable $n\mtfg{s}$ of CdO-3 and 4. This implies that the formation of virtual gap states at the surface is either complete before MgO deposition and the states are not affected by the passivation layer, or that the contact with MgO induces these states in the same way vacuum (or air) does. Both arguments seem plausible since there is no energetic overlap of extended electronic states of any MgO bands with the band gap region of CdO (type-I band alignment) due to the low electron affinity and large band gap of MgO (conduction band at $\approx E\mtfg{vac} - \SI{0.8}{\electronvolt}$ and valence band at $E\mtfg{vac} - \SI{8.6}{\electronvolt}$ \cite{Thomas1980}). The passivation layer does, however, significantly modify the electronic mobility of the surface system which may be attributed to scattering by localized states lying in the MgO/CdO interface gap.\\
It must be noted that the electron densities of the CdO layers, in particular CdO-3 and 4, are, to our knowledge, the lowest values ever reported in the literature, with the room temperature value of $n\mtfg{Hall} = \SI{5e18}{\per\cubic\centi\meter}$ being almost in the non-degenerate regime.

\subsection{Lateral transport properties of CdO/NiO heterostructures}
\begin{figure}
	\centering
	\includegraphics[width=\columnwidth]{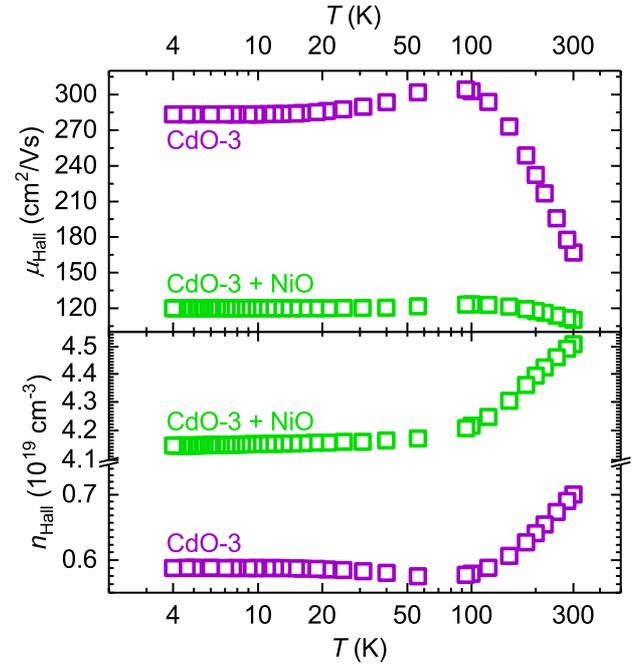}
	\caption{Hall carrier mobility and density for a CdO thin film before and after deposition of NiO on top}
	\label{fig:CdO+NiO-TdH}
\end{figure}

We next turn to CdO/NiO heterostructures. Fig.~\ref{fig:CdO+NiO-TdH} shows the temperature dependence of carrier mobility and density, as obtained from Hall effect measurements, of the sample CdO-3 before and after a \SI{50}{\nano\meter} thick NiO layer was deposited on top of it. In contrast to the MgO passivation, coating with NiO seems to radically alter the transport behavior of the CdO surface (the NiO film itself is not conductive and therefore does not contribute any conductivity or Hall voltage). The carrier density is enhanced by a factor of 6 and the mobility decreases by \SI{60}{\percent} at \SI{4}{\kelvin} as compared to bare CdO. This can be interpreted as indirect evidence of conductive interfacial states induced by the contact with NiO. 

\begin{figure}
	\centering
	\includegraphics[width=\columnwidth]{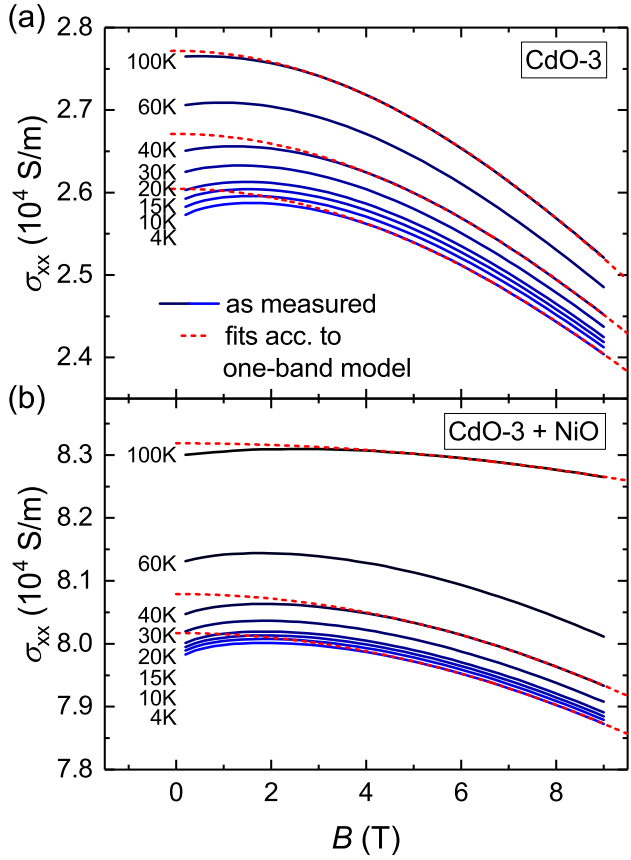}
	\caption{Diagonal elements $\sigma\mtfg{xx}$ of the magnetoconductivity tensor of (a) a bare CdO layer, (b) for a CdO/NiO heterostructure, for various temperatures. Red dashed lines are fits according to Eqn.~(\ref{eq:magnCond_lorentz}).}
	\label{fig:CdO+NiO_sigmaxx}
\end{figure}

\begin{figure}
	\centering
	\includegraphics[width=\columnwidth]{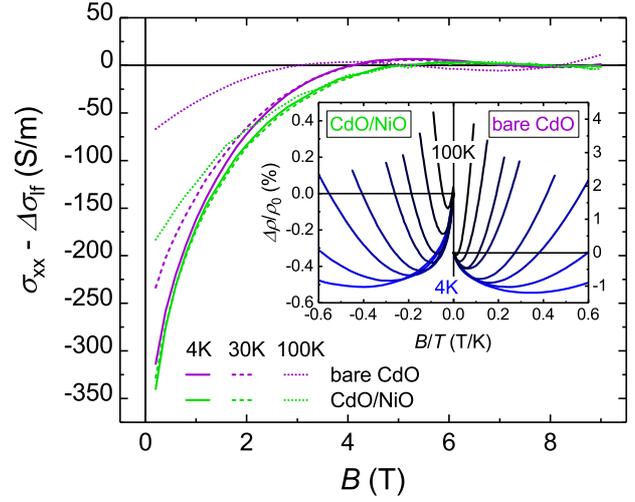}
	\caption{Magnetoconductivity $\sigma\mtfg{xx}$ for bare and NiO-coated CdO for selected temperatures, after correction for positive Lorentz-force contribution. Inset: plot of magnetoresistance vs. $B/T$ of both samples for all temperatures investigated.}
	\label{fig:corr_sigmaxx+MR_vs_B-T}
\end{figure}

Under certain conditions, Hall effect measurements on samples exhibiting parallel conduction by two electronic systems can be corrected in such a way that the two contributions can be distinguished from each other \cite{Look1997}. An important prerequisite is that the 2DEG observed at low tempertures possesses a constant carrier density and mobility across the whole temperature range investigated. Attempts to use this procedure on the data presented in this paper fails: after correction for the low-temperature degenerate system, the corrected carrier densities and mobilities show unphysical behaviour over a wide temperature range. The assumption of a temperature-independent conductivity of the 2D system must therefore be discarded. A reasonable explanation is that the electrons in the surface or interface layer are subjected to scattering processes with temperature-dependent scattering time constants such that the carriers' mobility changes with $T$. In order to study this, measurements of the magnetoconductivity were taken in the temperature range between \SI{4}{\kelvin} and \SI{100}{\kelvin} using magnetic fields between \SI{0}{\tesla} and \SI{9}{\tesla}, oriented perpendicular to the sample surface. The results are shown in Fig.~\ref{fig:CdO+NiO_sigmaxx} for the bare and the NiO-coated CdO film. The most striking features of these data is the pronounced increase of conductivity for magnetic fields between \SI{0}{\tesla} and $\approx \SI{2}{\tesla}$ (negative magnetoresistance, neg. MR). For both the bare and the NiO-coated CdO film, this effect can be observed to up to \SI{100}{\kelvin}. To make the low-field behaviour better visible, the decrease of $\sigma\mtfg{xx}$ at higher fields due to the Lorentz force was fitted with the standard formula:

\begin{equation}
	\Delta\sigma\mtfg{lf}(B) = \frac{\sigma_0}{1+(\mu B)^2} = \frac{e\mu n}{1+ (\mu B)^2}
	\label{eq:magnCond_lorentz}
\end{equation}

which agrees very well with the data above \SI{3}{\tesla}; the fits for selected temperatures are shown as red dashed lines in Fig.~\ref{fig:CdO+NiO_sigmaxx}. After subtracting the Lorentz force contribution, the remaining conductivity $\delta \sigma\mtfg{xx} = \sigma\mtfg{xx} - \Delta\sigma\mtfg{lf}$ is plotted in Fig.~\ref{fig:corr_sigmaxx+MR_vs_B-T}. It is clearly visible that the low-field contribution is the same for the bare and the NiO-coated CdO.

Negative MR in CdO has been reported by Wang \textit{et al.} \cite{Wang2016} and has therein been attributed to the weak localization (WL) effect. This phenomenon is related to constructive quantum interference of charge carrier wave functions that are localized around scattering centers in two dimensions. The effect lowers the conductivity of the material at small magnetic fields and temperatures. However, the present authors are convinced that the magnetic fields ($B > \SI{2}{\tesla}$) and temperatures ($T > \SI{100}{\kelvin}$) needed to destroy the effect in CdO -- both in our data and in that of Wang \textit{et al.} -- are too high to be due to WL. Goldstein, Grinshpan and Many have observed negative MR in surface accumulation layers on ZnO single crystals \cite{Goldstein1979}. A central observation of these authors is that the magnitude of the negative MR depended on the surface accumulation density and increased together with it. These authors have also developed a theory based on spin scattering, ie. interaction of electrons with ionized impurities via their respective magnetic moments. Since the cross-section of inelastic spin-dependent scattering is lowest when the spins of the scatterer and the charge carrier are parallel \cite{Matthew1982}, this mechanism becomes less dominant with increasing magnetic field. The theory in Ref.~\cite{Goldstein1979} also shows that the magnetoresistance 

\begin{equation}
\frac{\Delta\rho\mtfg{xx}}{\rho\mtfg{xx,0}} = \frac{\rho\mtfg{xx}(B)-\rho\mtfg{xx}(0)}{\rho\tfg{xx}(0)}
\end{equation}

is a universal function of $B/T$. The inset in Fig.~\ref{fig:corr_sigmaxx+MR_vs_B-T} displays the MR data of the bare and NiO-coated CdO film as a function of $B/T$, and it can be seen that the curves indeed follow a common line (for one specific sample) at low reduced fields before departing from it due to the Lorentz force contribution. Because spin scatttering only depends, in a first approximation, on the relative orientation of the spins, and not on the direction of current with respect to the magnetic field, the negative MR should be present also when the magnetic field is oriented in-plane (longitudinal magnetoresistance). The Lorentz force contribution, on the other hand, is zero when the field is  oriented parallel to the current flow. We have checked this on sample CdO-2 with an in-plane oriented magnetic field and found both facts to be fulfilled (data not included here) such that the interpretation of both contributions to the magnetoconductivity as presented here is well supported. It is probable that the temperature dependence of the spin-dependent scattering is what is causing a decreasing mobility of the surface system with rising temperature, causing the two-layer correction of the Hall data to fail.

We now turn back to the main assertion of the data in Fig.~\ref{fig:corr_sigmaxx+MR_vs_B-T}. The negative MR at low fields is a fingerprint of the CdO surface accumulation layer, and the MR magnitude is directly proportional to its sheet charge density. The fact that the negative MR magnitude does not change upon deposition of an NiO layer therefore implies that the CdO surface system remains unaltered. The sample's total sheet density, however, increases dramatically by a factor of 7 (see Fig~\ref{fig:CdO+NiO-TdH}). The CdO/NiO contact must therefore possess a second conductive channel with a sheet density of approximately \SI{9.7e14}{\per\square\centi\meter}. Because the negative character of the Hall effect is enhanced, this channel must be of $n$-type nature. Based on the simulations presented in Fig.~\ref{fig:NiO-CdO_band_alignment}, $p$-type behavior is expected for an enhancement layer on the NiO side; if such a layer exists, it is not detected in this experiment. The high surface electron density of the CdO layer can be assumed to efficiently screen the effects of charge redistrubution, such that a contact-induced enhancement layer in the CdO bulk (which would be $n$-type) is excluded. It is therefore concluded here that the CdO/NiO contact induces a conductive channel directly at the interface, as proposed by theory. Our results can be seen as a proof-of-concept for the type-III nature of the CdO/NiO heterocontact. It can be stated that the conductive interface channel seems to be stable up to room temperature, as $n\mtfg{Hall}$ retains its enhanced value across the whole temperature range investigated here. It shall also be noted that on the samples investigated in this work, we could not detect a chirality of the magnetotransport which would be indicative of helical edge states. We believe that in order to observe this, samples with well-defined edges have to be used (micro-Hall bar structures, for instance) because this would eliminate the influence of (trivially behaving) uncoated CdO at the edges of the sample. This issue must be addressed in future work.\\

\section{Conclusion}

We have investigated two types of heterojunctions comprising $p$-type NiO, both charaterized by large band offsets towards the second oxide. In the first case, ZnO was the $n$-type partner, forming a type-II band alignment with NiO. The electrical current through such a configuration is determined by the rate at which electrons and holes recombine at the interface. We have developed a simplified framework to model the parameters of the current-voltage relationship, and have demonstrated the critical impact of the interface quality in terms of defect population. A method to reduce the amount of interface defects introduced during NiO growth was given. The ZnO/NiO contacts can be employed as UV-converting solar cells with record efficiency; however, we presented evidence for a recombination process triggered by UV light, which we attribute to the previously identified recombination center Ni\tfg{Zn} that forms by in-diffusion of Ni into the topmost layers of ZnO.\\
The second structure investigated here is the CdO/NiO contact. We have shown that the interface (which is expected to have type-III character) harbors a conductive layer that is neither present in CdO nor in NiO alone, and that is detectable at least up to room temperature. We believe this to be a 2D carrier enhancement layer directly at the CdO/NiO interface which is apparently an $n$-type one, in accordance with theoretical expectations for such a type-III junction. This work can therefore serve as basis for the next step, which is proving the topological nature of the interfacial eletronic system.

\begin{acknowledgement}
This work was funded by the Deutsche Forschungsgemeinschaft (DFG) within the collaborative research center "SFB762: Functionality of Oxide Interfaces". The authors thank Jos\'{e} Barzola-Quiquia for fruitful discussions.
\end{acknowledgement}

%
\bibliographystyle{pss}
\bibliography{ref}

\end{document}